\documentclass[twocolumn,showpacs,amsmath,prl]{revtex4-1}

\usepackage{color}
\usepackage{graphicx}
\usepackage{epsfig,psfrag}
\usepackage{amsmath}
\usepackage{amssymb}
\usepackage{amsbsy}
\usepackage{amsthm}
\usepackage{amsfonts}
\usepackage{wasysym}
\usepackage{bbm}
\usepackage{tabularx}
\usepackage{euscript}
\usepackage{color}
\usepackage{enumerate}
\usepackage{amsfonts}
\usepackage{exscale}
\usepackage{bbold}
\usepackage{float}

\usepackage[colorlinks,citecolor=blue]{hyperref}

\newcommand{\Ref}[1]{Ref.~\cite{#1}}
\newcommand{\Eq}[1]{equation~(\ref{#1})}
\newcommand{\Fig}[1]{Fig.~\ref{#1}}

\newcommand{\ra}{\rightarrow}
\newcommand{\s}{{\sigma}}
\def\bea{\begin{eqnarray}}
\def\eea{\end{eqnarray}}

\def\Eq#1{Eq.~(\ref{#1})}
\def\Fig#1{Fig.~\ref{#1}}

\begin{document}

\title{What makes the $T_{\rm c}$ of monolayer FeSe on SrTiO$_3$ so high: a sign-problem-free quantum Monte Carlo study}

\author{Zi-Xiang Li$^{1}$,
Fa Wang$^{2,3}$, Hong Yao$^{1,3,\ast}$ \&
Dung-Hai Lee$^{4,5,\ast}$}
\affiliation{$^1$Institute for Advanced Study, Tsinghua University, Beijing 100084, China.\\
$^2$International Center for Quantum Materials, School of Physics, Peking University, Beijing 100871, China.\\
$^3$Collaborative Innovation Center of Quantum Matter, Beijing 100871, China.\\
$^4$Department of Physics, University of California, Berkeley, CA 94720, USA.\\
$^5$Materials Sciences Division, Lawrence Berkeley National Laboratory, Berkeley, CA 94720, USA.}

\begin{abstract}
\end{abstract}

\maketitle

{\bf Monolayer FeSe films grown on SrTiO$_3$ (STO) substrate show superconducting gap-opening temperatures ($T_{\rm c}$) which are almost an order of magnitude higher than those of the bulk FeSe and are highest among all known Fe-based superconductors.
Angle-resolved photoemission spectroscopy (ARPES) observed ``replica bands'' suggesting the importance of the interaction between FeSe electrons and STO phonons. These facts rejuvenated the quest for $T_{\rm c}$ enhancement mechanisms in iron-based, especially iron-chalcogenide, superconductors. Here, we perform the first numerically-exact sign-problem-free quantum Monte Carlo simulations to iron-based superconductors. We (i) study the electronic pairing mechanism intrinsic to heavily electron doped FeSe films, and (ii) examine the effects of electron-phonon interaction between FeSe and STO as well as nematic fluctuations on $T_{\rm c}$. Armed with these results, we return to the question ``what makes the $T_{\rm c}$ of monolayer FeSe on SrTiO$_3$ so high?'' in the conclusion and discussions.}\\
\maketitle


\noindent{{\bf 1. Introductions}}\\
\noindent
The strong Cooper pairing in monolayer FeSe film on SrTiO$_3$ substrate ((FeSe)$_1$/STO)  \cite{Xue} continues to attract a great deal of attentions (e.g. Refs. \cite{DLF,Liu,Tan,JFJia,He,YYWang,CTang-2015,Chen,Zhao, Niu, Miyata,Wen,Tang,JJ,DHL,YZhang}).
Recent developments in the study of FeSe-based high temperature superconductors clearly indicate there are at least two factors that are important to the enhancement of $T_{\rm c}$ from $8.9$ K (bulk FeSe) to about $75$ K in FeSe/BaTiO$_3$/SrTiO$_3$ \cite{DLF}. These factors are (1) heavy electron doping  \cite{Chen,Zhao, Niu, Miyata,Wen,Tang} and (2) the effects of the substrate \cite{JJ,DHL}.

\begin{figure}
\includegraphics[width=7.0cm]{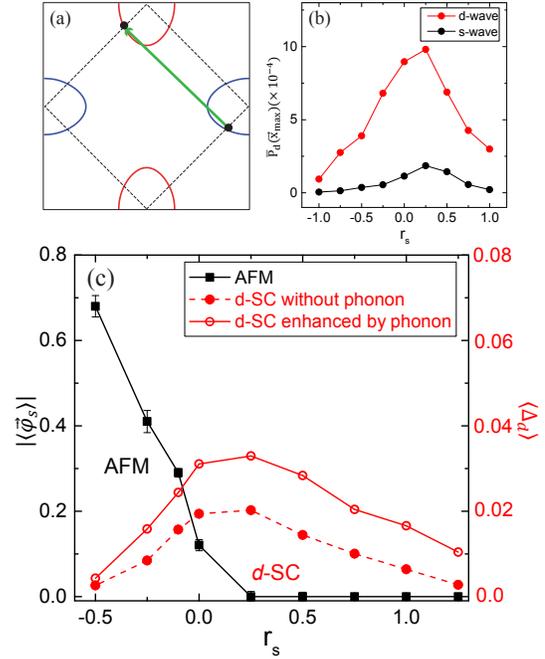}
\caption{\textbf{a} The Fermi surfaces of the bandstructure used in our simulations. The $J_1$-type AFM fluctuations can cause the inter-pocket scattering (green arrow). \textbf{b} The SC correlation in $s$- and $d$-wave pairing channels as a function of the parameter that controls the $J_1$-type spin fluctuations (the size $L=18$). \textbf{c} The phase diagram for the $J_1$-type spin fluctuations where the $d$-wave SC is substantially enhanced by the electron-phonon coupling (solid red curve) compared to the one without electron-phonon couplings (dashed red curve). Here we use the ground state expectation value of the AFM and SC order parameters as a measure of their ordering temperatures }
\label{phasediag}
\end{figure}

The first factor, namely heavy electron doping, shapes the fermiology into that best for the intrinsic electron pairing mechanism to act \cite{YZhang}. Concerning the intrinsic pairing mechanism there are two main candidates: the spin  \cite{Mazin2008, Kuroki2008, Chubukov2008, Seo2008, Wang2009,  Graser2009} and orbital \cite{Kontani} fluctuations mediated pairing. However, these proposals are based on approximations that are often not controlled in the presence of strong correlations. By now there are mounting experimental \cite{Basov, Tamai, Yi} and theoretical  \cite{Si,Yin,Georges} evidences that iron-based superconductors, in particular the iron-chalcogenide superconductors, are strongly correlated. Thus, a theoretical method free of uncontrolled approximations suitable for handling such situation is in high demand.

An experiment that sheds lots of light on the second factor, i.e., the effects of substrate,  is the ARPES result of \Ref{JJ}, which shows ``replica bands'' approximately 100 meV away from all low binding energy bands. Such phenomenon is explained in terms of ``phonon shake off'', and the phonons are identified with the longitudinal optical phonon branch of STO \cite{JJ,DHL}. This result suggests there is a strong coupling between the FeSe electrons and STO phonons. Moreover, it is conjectured that such coupling can substantially enhance the $T_{\rm c}$ intrinsic to heavily electron doped FeSe \cite{JJ,DHL}.

In the rest of the paper, we perform large-scale projector quantum Monte Carlo (QMC) \cite{Sorella-1989, White-1989,Assaad-2005} simulation (details are discussed in \Ref{Zi-Xiang-2015}). It turns out that the fermiology, namely the existence of two separate electron Fermi pockets, of (FeSe)$_1$/STO allows the simulation to be free of the fermion minus sign problem. This enables us to perform approximation-free unbiased study of the intrinsic electronic pairing mechanisms, namely, the antiferromagnetic (AFM) and antiferro-orbital (AFO) fluctuation mediated pairing. It also allows us to study the effects of electron-phonon interaction between FeSe and STO \cite{DHL} and nematic fluctuations \cite{Kivelson-2015, Berg-2015} on  $T_{\rm c}$.

A summary of our results is as follows. For the intrinsic pairing mechanisms we have studied two types of spin fluctuations and one type of orbital fluctuations. A commonality between these fluctuations  is that they all scatter electrons from one Fermi pocket to the other. (1) For spin fluctuations mimicking the nearest-neighbor AFM exchange interaction (the ``$J_1$-type'' spin fluctuation) the ground state exhibits {\it nodeless} $d$-wave superconducting (SC) long range order. (2) For spin fluctuations mimicking the next-nearest-neighbor AFM exchange interaction (the ``$J_2$-type''spin fluctuation) the ground state exhibits $s$-wave SC long range order. (3) The AFO fluctuations trigger $s$-wave pairing. For the enhancement mechanisms we have studied the small momentum transfer electron-phonon interactions and the nematic fluctuations. Our results clearly show (4) the small momentum transfer electron-phonon interaction significantly strengthens the Cooper pairing triggered by both spin and orbital fluctuations. (5) Similar to the electron-phonon interaction nematic fluctuations  also strengthen the Cooper pairing triggered by all three intrinsic mechanisms discussed above. A highlight of some of the main results is shown in \Fig{phasediag}.    \\

\noindent{{\bf 2. Sign-problem-free quantum Monte Carlo}}\\
\noindent The effective actions we consider are given in the Supplementary Information I$-$IV.
These actions consist of three parts: (1) the bandstructure of electrons, (2) various fluctuating Bose fields, and (3) the ``Yukawa'' coupling between the Bose fields and electrons.
The bandstructure is chosen to mimic the Fermi surfaces of (FeSe)$_1$/STO as shown in \Fig{phasediag}a. We use the one-iron Brillouin zone because it has been shown experimentally that when folded to the corners of the two-iron Brillouin zone the electron pockets show negligible hybridization at their crossings \cite{Yan}.

For intrinsic pairing mechanisms the Bose fields we studied include $\vec{\varphi}_{\rm s}$ and $\varphi_{\rm o}$ associated with the spin and orbital fluctuations respectively.
These fields scatter electrons between the Fermi pockets as shown by the green arrow in \Fig{phasediag}a.
For the pairing enhancement mechanisms, we studied
STO phonons and nematic fluctuations. The Bose fields associated with them are $\varphi_{\rm ph}$ and $\varphi_{\rm n}$, they cause small momentum transfer ({\it i.e.} intra-pocket) scattering of the FeSe electrons. The reason we only consider small momentum phonon scattering is due to the forward-focusing nature of the electron-phonon interaction deduced from Ref. [\onlinecite{JJ}].
In Eqs.~(S2), Eq.~(S5) and  Eq.~(S9), the parameters $r_{\rm s,o,n}$ tune $\vec{\varphi}_{\rm s}$, $\varphi_{\rm o}$, $\varphi_{\rm n}$ across their respective quantum phase transitions. Large negative values correspond to strongly ordered phase and large positive values correspond to the strongly disordered phase. The parameter $r_{\rm ph}$ in Eq.~(S7) controls the optical phonon frequency at $\vec{q}=0$. Remarkably, in all cases our QMC calculation has no minus sign problem \cite{Wu-05,Berg-12,Li-15} (see Supplemental Material VI). The QMC simulation is carried out on a square lattice with $N=L\times L$ sites using periodic boundary conditions. In the following we present the simulation results.\\

\begin{figure}[t]
\includegraphics[width=8.5cm]{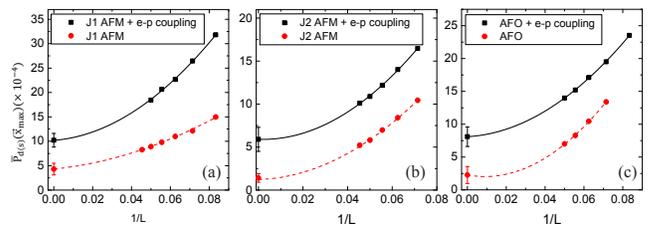}
\caption{
The enhancement of SC correlation by the small momentum transfer electron-phonon interaction. In each panel the SC correlation in the dominant pairing channel, $\overline{P}_{d}(L/2,L/2)$, is plot as a function of $1/L$ with and without the electron-phonon interactions. Panels (\textbf{a})$-$(\textbf{c}) are for  $J_1$-type spin (\textbf{a}), $J_2$-type spin (\textbf{b}), and AFO (\textbf{c}) fluctuations, respectively
}
\label{e-ph}
\end{figure}

\noindent{{\bf 3. The spin fluctuation mediated pairing}}\\
\noindent {\bf (1) The $J_1$-type spin fluctuation:} The effective action is given by Eqs.~(S1)--(S3). The reason we refer to it as the $J_1$-type spin fluctuation is because  integrating out $\vec{\varphi}_{\rm s}$  generates an AFM exchange interaction whose momentum space coupling constant has the same sign has that of the nearest-neighbor ($J_1$) AFM exchange interaction. From the Binder cumulant \cite{binder} of the AFM order parameter (not shown), we estimate the AFM quantum critical point $r_{\rm s,c}$ to lie in the range of $(0,0.25)$.
To study superconductivity we compute the equal-time pair-pair correlation function $\overline{P}_{s/d}(\vec x_{\rm max})$ (see Eqs.~(S11)--(S13)). Here $s/d$ denotes $s$-wave (same sign on the two electron pockets) and (nodeless) $d$-wave (opposite sign on the two electron pockets) pairing, respectively. $\vec x_\textrm{\rm max}=(L/2,L/2)$ is the maximum separation between the two pair fields in a system of size $L$. In \Fig{phasediag}b, we plot $\overline{P}_{s/d}(L/2,L/2)$ for $L=18$ as a function of $r_{\rm s}$. Clearly superconductivity is enhanced near the magnetic quantum critical point. Moreover, the $d$-wave pairing is favored over the $s$-wave  \cite{Dunghai-2013}.

In \Fig{e-ph}a (red curve), we show the size-dependence of $\overline{P}_{d}(L/2,L/2)$ at $r_{\rm s}=0.25$ for $L = 12, 14, 16, 18, 20, 22$ (the red points). The red curve is the best fit using a second order polynomial in $1/L$. This allows us to extrapolate to $L\ra\infty$ to obtain $\overline{P}_{d}(L \ra\infty) = (4.3 \pm 1.1) \times 10^{-4}$.
This establishes the fact that the ground state possesses nodeless $d$-wave superconducting long-range order!\\

\begin{figure}[t]
\includegraphics[width=8.6cm]{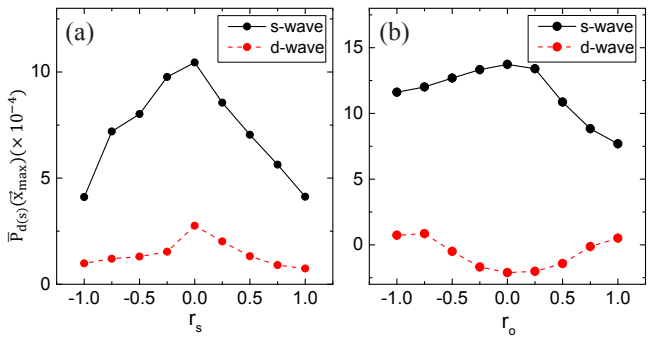}
\caption{The SC correlation, $\overline{P}(L/2,L/2)$, in the $s$- and $d$-wave pairing channels for $L=14$ triggered by the (\textbf{a}) $J_2$-type spin and (\textbf{b}) AFO fluctuations. For both cases, the $s$-wave pairing is favored over the $d$-wave }
\label{intrinsic}
\end{figure}

\noindent {\bf (2) The $J_2$-type spin fluctuation:} The effective action is given by Eqs.~(S1),(S2) and (S4). In this case, integrating out the spin boson $\vec{\varphi}_{\rm s}$  generates an AFM exchange interaction whose momentum space coupling constant has the same sign has that of the next nearest neighbor ($J_2$) AFM exchange interaction. From the Binder cumulant  (not shown here) we deduce the quantum critical point to be situated within $ 0.0 \le r_{\rm s,c}\le 0.25$. In \Fig{intrinsic}a, we plot $\overline{P}_{s/d}(L/2,L/2)$ for $L=14$ as a function of $r_{\rm s}$. Here  $s$-wave superconductivity is enhanced near the magnetic quantum critical point.

In \Fig{e-ph}b (red curve), we study the size-dependence of $\overline{P}_{d}(L/2,L/2)$ at $r_{\rm s}=0.25$ for $L = 12, 14, 16, 18, 20, 22$ (the red points). The red curve is the best fit using a second order polynomial in $1/L$.
This allows us to extrapolate to $L\ra\infty$ to obtain $\overline{P}_{d}(L \ra\infty) = (1.4\pm0.5)\times 10^{-4}$. This establishes the fact that the ground state possesses nodeless $s$-wave superconducting long-range order. \\

\noindent{{\bf 4. The AFO fluctuation mediated pairing}}\\
\noindent In this section, we study the effects of AFO fluctuation ($\varphi_o$) on superconductivity.
The effective action is given by Eqs.~(S1), (S5) and (S6).
Like the AFM fields the AFO field also scatters electrons between the two Fermi pockets (the green arrow in \Fig{phasediag}a).

From the Binder cumulant associated with the AFO order parameter (not shown here) we deduce the AFO quantum critical point to be situated within $ 0.0 \le r_{\rm o,c}\le 0.25$.
In \Fig{intrinsic}b,  we plot $\overline{P}_{s/d}(L/2,L/2)$ for $L=14$ as a function of $r_{\rm o}$.
Clearly $s$-wave superconducting correlation is favored over the $d$-wave, and it is peaked near the AFO quantum critical point. In \Fig{e-ph}c (red curve) we study the size-dependence of $\overline{P}_{d}(L/2,L/2)$ at $r_{\rm o}=0.25$ for $L = 12, 14, 16,18,20,22$ (the red points). The red curve is the best fit using a second order polynomial in $1/L$. This allows us to extrapolate to $L\ra\infty$ to obtain $\overline{P}_{s}(L \ra\infty) = (2.1 \pm 0.9)\times 10^{-4}$. This indicates that the ground state possesses $s$-wave superconducting long-range order.\\

\noindent{{\bf 5. The pairing enhancement due to STO phonons}}\\
\noindent Motivated by \Ref{JJ}, here we study the effect of small momentum transfer electron-phonon coupling on the superconductivity triggered by pure AFM and AFO fluctuations.
This is done by adding the coupling to $\varphi_{\rm ph}$ (see Eqs.~(S7), (S8)).
The parameter $r_{\rm ph}$ that controls the phonon frequency is fixed at 0.5. The strength of the electron-phonon coupling is controlled by  $\lambda_{\rm ph}$. The value of $\lambda_{\rm ph}$ is chosen so that the dimensionless strength of the phonon mediated attraction $\lambda = \frac{\lambda_{\rm ph}^2}{r_{\rm ph} W}=0.6$. Here $W$ is the electron band width.  This value is similar to the estimate given in \Ref{JJ}. In the following we fix the parameter $r_{\rm s,o}$ at 0.25.
In \Fig{e-ph} we compare the size dependence of the superconducting correlation function in the dominant pairing channels with (black curve) and without (red curve) phonons. Clearly the SC order is enhanced by the electron-phonon interaction regardless of the intrinsic pairing mechanisms.

The phase diagram in \Fig{phasediag}c is constructed from the extrapolated value of the AFM and SC order parameters from finite-size analysis for each $r_{\rm s}$. The plot is for the $J_1$-type spin fluctuation, however we expect a similar plot holds for $J_2$-type spin and AFO fluctuations as well. In the phase diagram, we use the ground state expectation value of the AFM and SC
order parameters as a measure of their ordering temperatures. It is clear that the SC ordering temperature $T_{\rm c}$ is enhanced by the electron-phonon couplings for all value of $r_{\rm s}$. Remarkably, the $T_{\rm c}$ enhancement by phonons is largest around the AFM quantum critical point.

In Supplementary Materials VII, we study the enhancement of the superconducting order parameter due to $J_1$-type spin and AFO fluctuations as a function of the dimensionless phonon-mediated attraction strength $\lambda$. Apparently, the enhancement of superconductivity peaks at $\lambda=1.5$ for the $J_1$-type spin fluctuation triggered $d$-wave pairing. For the
AFO induced $s$-wave pairing the pair-pair correlation increases monotonously with the electron-phonon coupling strength up to $\lambda=2.2$. \\

\begin{figure}[t]
\includegraphics[width=8.5cm]{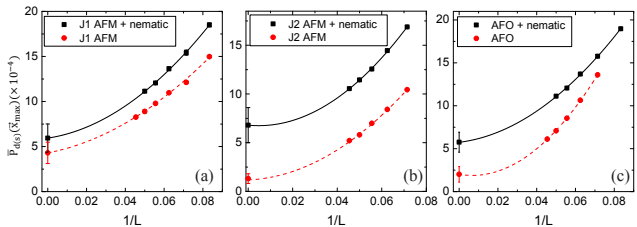}
\caption{
The enhancement of SC correlation by nematic fluctuations. In each panel, the SC correlation in the dominant pairing channel, $\overline{P}_{d}(L/2,L/2)$, is plot as a function of $1/L$ with and without the electron phonon interaction. Panels (\textbf{a})$-$(\textbf{c}) are for $J_1$-type  spin (\textbf{a}) $J_2$-type spin (\textbf{b}) and AFO (\textbf{c}) fluctuations, respectively
}
\label{nematicenh}
\end{figure}

\noindent{{\bf 6. The pairing enhancement by nematic fluctuations}}\\
\noindent In view of the possibility that nematic fluctuation can be substantial in heavily electron-doped FeSe films \cite{YZhang}, here we study the effects of nematic fluctuations on superconductivity. The effective action is given by Eqs.~(S1),(S9) and (S10).

In \Fig{nematicenh}, we compare the size dependence of the superconducting correlation function in the dominant pairing channels with (black curve) and without (red curve) nematic fluctuations. Like the electron-phonon interaction, the SC order is enhanced by the nematic fluctuations for all intrinsic pairing mechanisms considered.\\

\noindent{{\bf 7. Conclusion and discussions}}\\
\noindent A definitive answer to ``why $T_{\rm c}$ is so high in monolayer FeSe on SrTiO$_3$?''  requires one to (1) determine the intrinsic pairing mechanisms which is primarily responsible to Cooper pairing in heavily electron doped FeSe-based high temperature superconductors, and (2) pin down the effects of substrate.

Regarding (2), our results show that small momentum transfer electron-phonon scattering enhances
superconductivity regardless of whether it is triggered by the spin or orbital fluctuations, hence lend support to the phonon enhancement mechanism discussed in Refs. \cite{JJ,DHL}. However our result holds for all phonons that scatter the FeSe electron with small momentum transfer. It does not allow us to conclude that the particular branch of high frequency phonon which caused the replica bands in \Ref{JJ} is {\it solely} responsible for the $T_{\rm c}$ enhancement. In particular it does not rule out the importance of other lower frequency polar phonons.

Regarding (1), our results do not allow us to answer whether spin or orbital fluctuation is main intrinsic pairing mechanism in heavily electron doped FeSe films. However we can confidently predict the pairing symmetry associated with each pairing mechanism. In particular if the pairing symmetry turns out to be $s$-wave it can come from several different mechanisms: $J_2$-type spin fluctuation or  anti-ferro orbital fluctuation, or the combination of them with nematic fluctuation. However if the pairing symmetry is $d$-wave our result uniquely pins down the $J_1$-type spin fluctuation as the driving force.

Experimentally the pairing symmetry of (FeSe)$_1$/STO is still an open question. However we would like to list a number of circumstantial evidence that the pairing symmetry might be $d$-wave. The first is the existence of neutron resonance {\it below twice the superconducting gap} in materials with similar fermiology \cite{Park,Friemel}, and the fact that the momentum locations of the resonance are consistent with inter-pocket scattering. The second is a recent high resolution ARPES study of the superconducting gap anisotropy of  (FeSe)$_1$/STO \cite{Yan}. It observes four minima in the superconducting gap at the momentum locations corresponding to the crossing of the two Fermi pockets (in the two iron Brillouin zone). This  can be interpreted as the result due to weak inter-pocket hybridization on a {\it nodeless} $d$-wave gap \cite{Mazin}. Moreover the weakness of the inter-pocket hybridization is evidenced by the lack of splitting at the Fermi pocket crossings in the normal state.

Although there is no direct evidence of strong nematic fluctuation in (FeSe)$_1$/STO we can not rule out that it does play a partial role in the pairing of heavily electron doped FeSe films. By itself nematic fluctuation will not discriminate between $s$- and $d$-wave pairings. However when coupled with the spin or orbital fluctuations it can significantly enhance the pairing strength favored by each of them.     \\

{\noindent\bf Acknowledgments} \\
\noindent We would like to thank the National Supercomputer Center in Guangzhou for computational support.
ZXL and HY were supported in part by the National Thousand Young-Talents Program and the National Natural Science Foundation of China (11474175).
FW was supported by the National Natural Science Foundation of China(11374018).
DHL was  supported  by  the  U.S.  Department  of  Energy,  Office  of  Science,  Basic Energy Sciences, Materials Sciences and Engineering Division, grant DE-AC02-05CH11231.

{\it Note added}.-- After our work was posted on arXiv.org, another sign-problem-free QMC on FeSe appeared \cite{Ashvin-2015}. However the focus of  Ref.\cite{Ashvin-2015} is quite different from ours.\\

{\noindent\bf Conflict of interest} \\
The authors declare that they have no conflict of interest.\\

\noindent $^\ast$yaohong@tsinghua.edu.cn,~dunghai@berkeley.edu.

\begin{widetext}

\section{Supplemental Materials}

\renewcommand{\theequation}{S\arabic{equation}}
\setcounter{equation}{0}
\renewcommand{\thefigure}{S\arabic{figure}}
\setcounter{figure}{0}
\renewcommand{\thetable}{S\arabic{table}}
\setcounter{table}{0}

\subsection{I. The effective action for the $J_1$ and $J_2$ types of spin fluctuations}
The effective action, based on a two band model describing band structure of single-layer (FeSe)$_1$/STO, is given by $S=S_{\rm F} +S_{\rm s}$ where $S_{\rm s}=S_{\rm B}+S_{\rm FB}$ and
\bea
S_{\rm F} &=&\int_0^\beta {\rm d}\tau\sum_{jk,\alpha=x,y} \psi_{j\alpha}^{\dagger}\big[(\partial_\tau -\mu)\delta_{jk}-t_{jk,\alpha}\big]\psi_{k\alpha}, \\
S_{\rm B}&=& \int_0^\beta {\rm d}\tau\Big\{ \frac{1}{2}\sum_j \frac{1}{c_{\rm s}^2}|\partial_\tau \vec{\varphi}_{{\rm s}, j}|^2 + \frac{1}{2}\sum_{\langle jk\rangle}|\vec{\varphi}_{{\rm s}, j}-\vec{\varphi}_{s,k}|^2
  +\sum_j\Big[\frac{r_{\rm s}}{2}|\vec{\varphi}_{{\rm s}, j}|^2+\frac{u_{\rm s}}{4}\left(|\vec{\varphi}_{{\rm s}, j}|^2\right)^2\Big]\Big\},
\label{action}
\eea
where
\bea
S_{\rm FB} &=&\lambda_{\rm s}\int_0^\beta {\rm d}\tau\sum_j (-1)^j \Big[\psi^{\dagger}_{jx}(\vec{\sigma}\cdot\vec{\varphi}_{{\rm s}, j})\psi_{jy}+h.c.\Big],
\eea
for $J_1$-type of spin fluctuations. If the spin fluctuation is $J_2$ type
\bea
S_{\rm FB}=i~\lambda_{\rm s}\int_0^\beta {\rm d}\tau\sum_j (-1)^j \Big[\psi^{\dagger}_{jx}(\vec{\sigma}\cdot\vec{\varphi}_{{\rm s}, j})\psi_{jy}+h.c.\Big].
\label{action2}
\eea
In the above equations, $j,k$ labels the sites of a square lattice, $\alpha=x,y$ labels the two orbitals (which transform into each other under the 90 $^\circ$ rotation)
 from which the red and blue Fermi surfaces in Fig. 1a of the main text  
are derived from, $\tau$ denotes the imaginary time and $\beta$ is the inverse temperature.
In \Eq{action}, $\vec{\varphi}_{\rm s}$ is the AFM collective mode and the operator $\psi_{i\alpha}$ is a spinor operator which annihilates an electron in orbital $\alpha$ and on site $i$.
The three $\vec{\sigma}$ are the spin Pauli matrices. It is important to note that in \Eq{action2} the fermion-boson coupling has an extra factor $i$.

The parameters in this effective action include 
$r_{\rm s}$ which tunes the system across the AFM phase transition, $c_{\rm s}$ is the spin-wave velocity and $u_{\rm s}$ is the self-interactions of $\vec{\varphi}_{\rm s}$.
$\lambda_{\rm s}$ is the ``Yukawa'' coupling between electrons and AFM order parameter.
In our computation, we fix $c_{\rm s} = u_{\rm s} = \lambda_{\rm s}  = 1.0$ and vary the value of $r_{\rm s}$ to control the severity of AFM fluctuation. In the fermion action the hopping integral $t_{ij}$ is chosen to be among nearest neighbor sites and equal to $t_{\parallel}= 1.0$ for $x(y)$-orbital along $x(y)$ direction and $t_\perp = -0.5$ for $y(x)$ orbital along $x(y)$ direction.
We fix the occupancy in our computation to be $0.1$ such that the Fermi surface is shown in Fig. 1a of the main text.

\subsection{II. The effective action for antiferro-orbital fluctuations}
The AFO effective action is given by $S=S_{\rm F} +S_{\rm o}$ where $S_{\rm F}$ is the same as in Eq.(S1),   $S_{\rm o}=S_{\rm B}+S_{\rm FB}$ and
\bea
S_{\rm B}&=& \int_0^\beta {\rm d}\tau\Big\{ \frac{1}{2}\sum_i \frac{1}{c_{\rm o}^2}|\partial_\tau \varphi_{{\rm o},i}|^2 + \frac{1}{2}\sum_{\langle ij\rangle}|\varphi_{{\rm o},i}-\varphi_{{\rm o},j}|^2
  +\sum_i\Big[\frac{r_{\rm o}}{2}(\varphi_{{\rm o},i})^2+\frac{u_{\rm o}}{4}\left(\varphi_{{\rm o},i}\right)^4\Big]\Big\},\\
S_{\rm FB} &=&\lambda_{\rm o}\int_0^\beta {\rm d}\tau\sum_i (-1)^i \Big[\varphi_{{\rm o},i} \psi^{\dagger}_{ix}\sigma_0\psi_{iy}+h.c.\Big].
\label{actiono}
\eea
In \Eq{actiono}, $\varphi_{\rm o}$ is the AFO order parameter and  $\sigma_0$ is the identity matrix in the spin space.

The parameters in this effective action include the orbital wave velocity $c_{\rm o}$, and $r_{\rm o}$ which tunes the system across the AFO phase transition, and $u_{\rm o}$ is the self-interactions of the $\varphi_{\rm o}$ field.
$\lambda_{\rm o}$ is the Yukawa coupling between electrons and AFO order parameter.
In our computation, we fix $c_{\rm o} = u_{\rm o} = \lambda_{\rm o}  = 1.0$ and vary the value of $r_{\rm o}$ to control the severity of AFO fluctuation.

\subsection{III. The electron-phonon effective action}
The electron-phonon effective action is given by  $S=S_{\rm F} +S_{\rm ph}$ where $S_{\rm ph}=S_{\rm B}+S_{\rm FB}$ and
\bea
S_{B} &=&\int_0^\beta {\rm d}\tau\Big\{ \frac{1}{2}\sum_i \frac{1}{c_{\rm ph}^2}|\partial_\tau \varphi_{{\rm ph},i}|^2 + \frac{1}{2}\sum_{\langle ij\rangle}|\varphi_{{\rm ph},i}-\varphi_{{\rm ph},j}|^2
  +\sum_i\Big[\frac{r_{\rm ph}}{2}|\varphi_{{\rm ph},i}|^2\Big]\Big\}, \\
S_{\rm FB} &=& \lambda_{\rm ep}\int_0^\beta {\rm d}\tau\sum_i \varphi_{{\rm ph},i} \Big[\psi^{\dagger}_{ix}\s_0\psi_{ix}+\psi^{\dagger}_{iy}\s_0\psi_{iy}\Big].
\label{eph}
\eea
Here $\varphi_{\rm ph}$ is the phonon field, $r_{\rm ph}$ is the frequency of the optical phonon at $\vec{q}=0$ and $c_{\rm ph}$ is velocity of phonon. We fix parameters $r_{\rm ph} = 0.5, c_{\rm ph} = 1.0$ and vary the value of $\lambda_{\rm ep}$ to tune the strength of electron-phonon coupling.

\subsection{IV. The effective action for nematic fluctuations}
In \Fig{fs}, we show how does the nematic order parameter distort the Fermi surfaces.
The nematic effective action is given by $S=S_{\rm F} +S_{\rm n}$ where $S_{\rm n}=S_{\rm B}+S_{\rm FB}$ and
\bea
S_{\rm B}&=&\int_0^\beta {\rm d}\tau\Big\{ \frac{1}{2}\sum_i \frac{1}{c_{\rm n}^2}|\partial_\tau \varphi_{{\rm n},i}|^2 + \frac{1}{2}\sum_{\langle ij\rangle}|\varphi_{{\rm n},i}-\varphi_{{\rm n},j}|^2
  +\sum_i\Big[\frac{r_{\rm n}}{2}|\varphi_{{\rm n},i}|^2+\frac{u_{\rm n}}{4}\varphi_{{\rm n},i}^4\Big]\Big\},\\
S_{\rm FB} &=&\lambda_{\rm n}\int_0^\beta {\rm d}\tau\sum_i \varphi_{{\rm n},i} \Big[\psi^{\dagger}_{ix}\s_0\psi_{ix}-\psi^{\dagger}_{iy}\s_0\psi_{iy}\Big].
\label{actionn}
\eea
In \Eq{actionn}, $\varphi_{\rm n}$ is the nematic order parameter.
The parameters in this effective action include the velocity $c_{\rm n}$, and $r_{\rm n}$ which tunes the system across the nematic phase transition, and $u_{\rm n}$ is the self-interactions of the $\varphi_{\rm n}$ field.
$\lambda_{\rm n}$ is the Yukawa coupling between electrons and nematic order parameter.
In our computation, we fix $c_{\rm n} = u_{\rm n} = \lambda_{\rm n}  = 1.0$ and vary the value of $r_{\rm n}$ to control the severity of nematic fluctuation.
\begin{figure}[t]
\includegraphics[width=6.0cm]{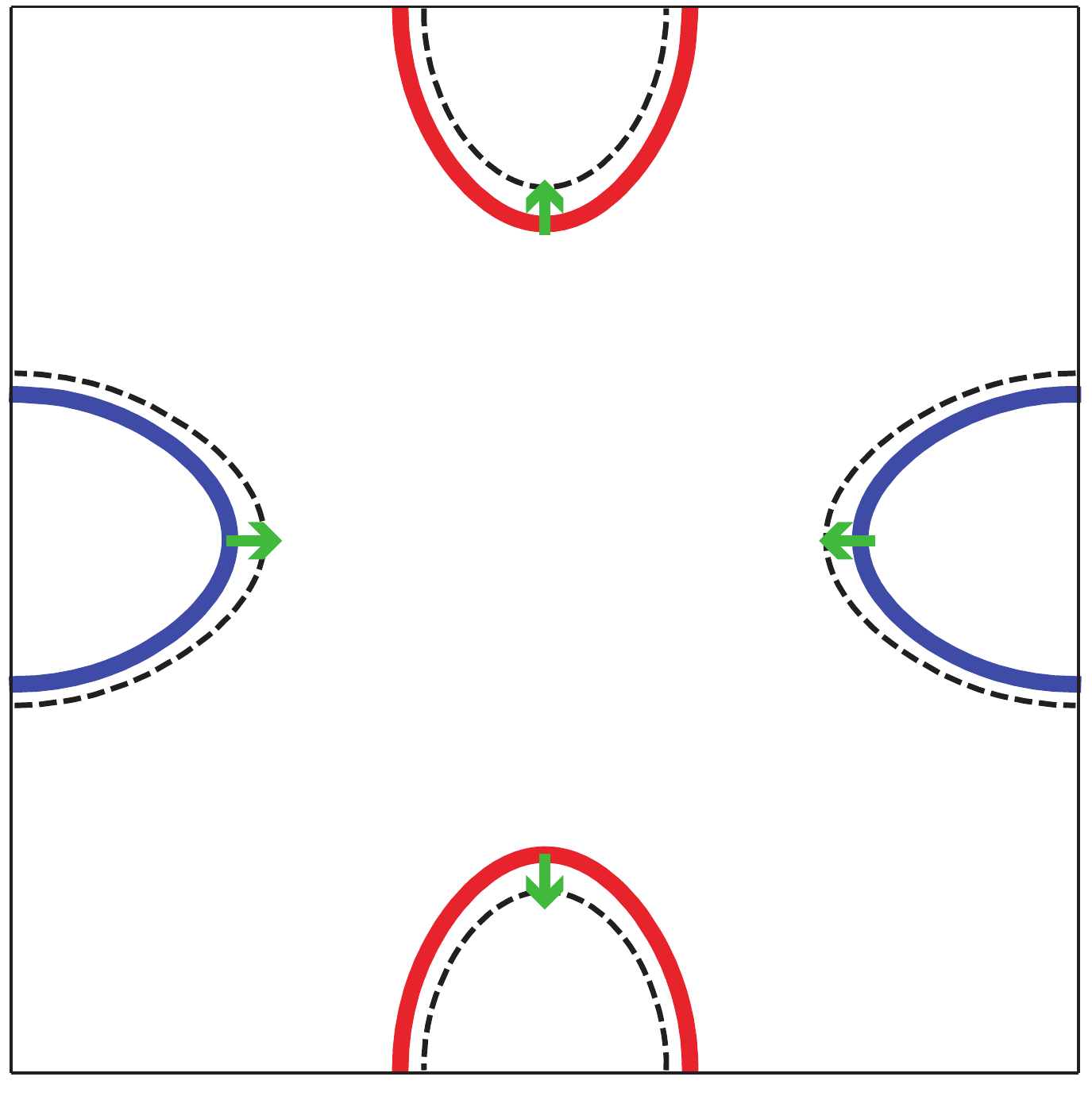}
\caption{
 Dashed lines represent the nematically distorted Fermi surfaces
}
\label{fs}
\end{figure}

\subsection{V. The superconducting pair correlation function}

To investigate superconductivity we calculate the equal time pair-pair correlation functions
\bea
P_{s/d}(\vec r_i)= \langle\Delta_{s/d}(\vec r_i)\Delta_{s/d}^{\dagger}(\vec 0)\rangle,
\eea
where
\bea
\Delta_{s/d}(\vec r_i)= \psi^{\rm T}_{ix}(i\s_y)\psi_{ix}\pm\psi^{\rm T}_{iy}(i\s_y)\psi_{iy},
\eea
are the $s$ (+ sign) and $d$ ($-$ sign) wave Cooper pair operators, respectively.
To determine whether there is long-range order we put $\vec{r}_i$ to the maximum separation $\vec{x}_{\rm max}$ of the pair fields (for a system with linear dimension $L$ the value of $\vec{x}_{\rm max}$ is $(L/2,L/2).$
Moreover, in order to minimize statistical errors, we average the correlation function over 25 sites around $\vec x_\textrm{max}$. 
Thus the actual pair correlation we study is
\bea
\overline{P}_{d(s)}(\vec x_\textrm{max})=\frac{1}{25}\sum_{n,m=0,\pm 1,\pm 2} P_{d(s)}(\vec x_\textrm{max}+n\hat x + m\hat y).
\label{SCC}
\eea

\begin{figure*}[b]
\includegraphics[width=15.0cm]{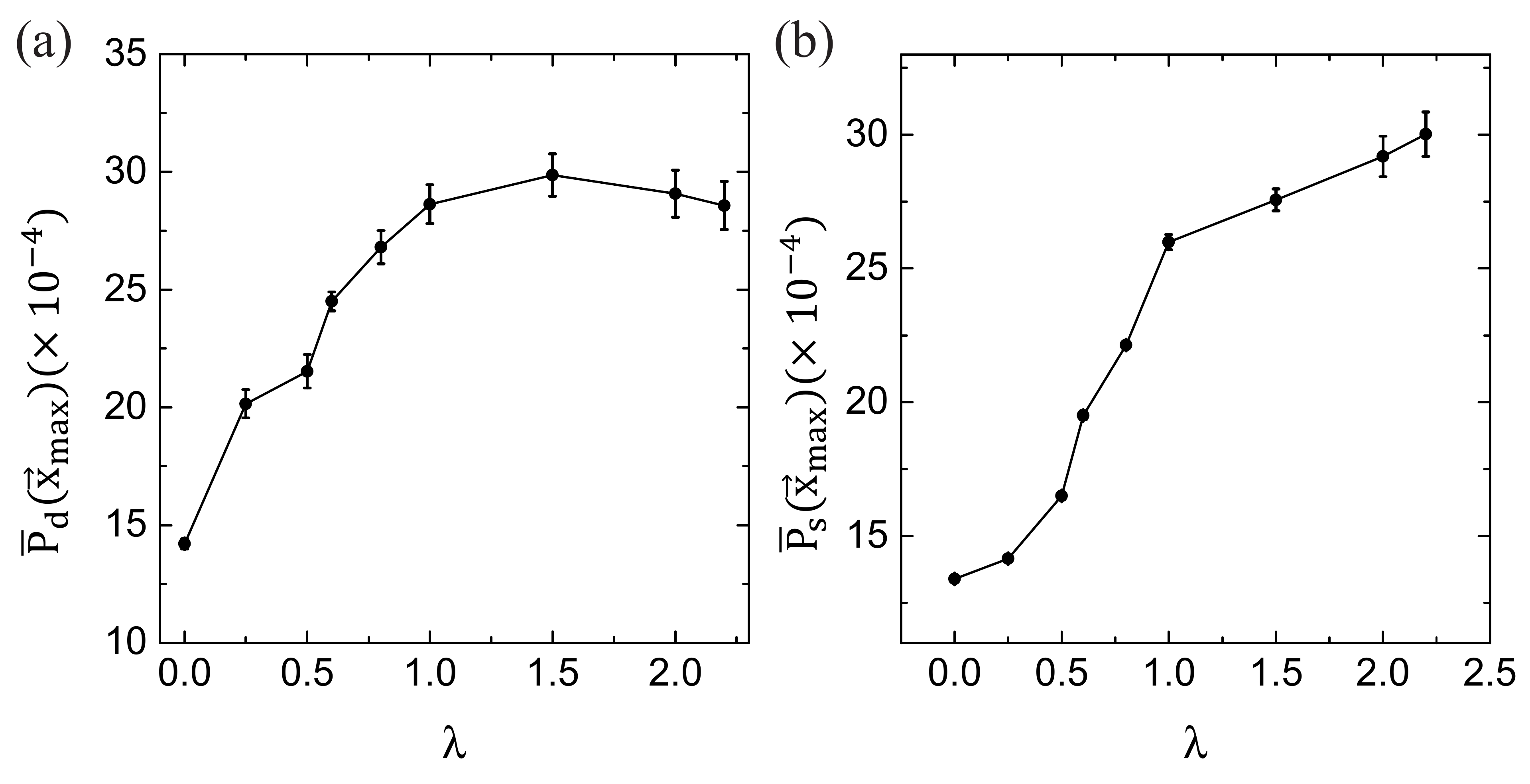}
\caption{
Enhancement of pairing by electron-phonon coupling.
\textbf{a} The $J_1$-type spin fluctuation triggered $\overline{P}_{d}(L/2,L/2)$ as a function of $\lambda$ for $L = 14$.
\textbf{b} The AFO fluctuation triggered $\overline{P}_{s}(L/2,L/2)$  as a function of $\lambda$ for $L = 14$
}
\label{phonon}
\end{figure*}

\subsection{VI. Effective actions that are amenable to sign-problem-free QMC simulation}

The actions that are amenable to sign-problem-free QMC simulations are any mixture of $S_{\rm F}+c_1 S_{s}+c_2 S_{\rm ph}+c_3 S_{\rm n}$ and $S_{\rm F}+c_1 S_{o}+c_2 S_{\rm ph}+c_3 S_{\rm n}$ where $c_{1,2,3}=0,1$. Note, however, for $S_{\rm s}$ we can use either $J_1$ or $J_2$ type spin fluctuation actions but not both.

Aside from the square lattice spatial symmetries the action  $S_{\rm F}+c_1 S_{s}+c_2 S_{\rm ph}+c_3 S_{\rm n}$ is invariant under the anti-unitary transformation $U = \tau_z(i\sigma_y) K$ and the action $S_{\rm F}+c_1 S_{o}+c_2 S_{\rm ph}+c_3 S_{\rm n}$ is invariant under $U^\prime = i\sigma_y K$.
Here $\tau_z$ is the third Pauli matrix acting in orbital ($x,y$) space and $K$ denotes complex conjugation.
It can be shown that because of these symmetries, the fermion determinant for arbitrary Bose fields configuration is positive hence the QMC simulation is free of minus-sign.
This enables us to perform large-scale projector QMC simulation.
The anti-unitary symmetry $U = \tau_z(i\sigma_y) K$ is the same as that in \Ref{Berg-12S}.

\subsection{VII. The enhancement of superconductivity triggered by the $J_1$ spin and AFO fluctuations as a function of the electorn-phonon coupling}

In this section, we examine the enhancement of the SC order parameters triggered by the $J_1$ type spin and AFO fluctuations as a function of the dimensionless electron-phonon coupling strength $\lambda$.

In \Fig{phonon}a and b, we plot the enhancement of spin fluctuation induced $\overline{P}_{d}(L/2,L/2)$  and orbital fluctuation induced $\overline{P}_{s}(L/2,L/2)$ as a function of $\lambda$ for $L = 14$.
Apparently the enhancement of superconductivity peaks at $\lambda\sim 1.5$ for the $J_1$-type spin fluctuation triggered $d$-wave pairing.
For the AFO induced $s$-wave pairing the  pair correlation increases monotonously with the amplitude of electron-phonon coupling up to the maximum value of  $\lambda$ we studied (=2.2).\\

\end{widetext}

\end{document}